\documentclass[runningheads]{llncs}
\usepackage[T1]{fontenc}
\usepackage{graphicx}
\usepackage[hidelinks]{hyperref}
\usepackage{dcolumn}
\usepackage{color}

\newcommand\blfootnote[1]{%
  \begingroup
  \renewcommand\thefootnote{}\footnote{#1}%
  \addtocounter{footnote}{-1}%
  \endgroup
}
\begin{document}
\title{Estimation of the Impact of COVID-19 Pandemic Lockdowns on Breast Cancer Deaths and Costs in Poland using Markovian Monte Carlo Simulation}
\titlerunning{Lockdowns effects on Breast Cancer Deaths and Costs}
\author{Magdalena Dul\inst{1,2} \and
Michal K. Grzeszczyk\inst{1}\orcidID{0000-0002-5304-1020} \and
Ewelina Nojszewska\inst{2}\orcidID{0000-0003-3176-0240} \and Arkadiusz Sitek \inst{3}\orcidID{0000-0002-0677-4002} }

\authorrunning{M. Dul {\em et al.}}
\institute{Sano Centre for Computational Medicine, Cracow, Poland 
\and Warsaw School of Economics, Warsaw, Poland
\and Massachusetts General Hospital, Harvard Medical School, Boston, MA, USA
}
\maketitle              % typeset the header of the contribution

\blfootnote{M. Dul and M. K. Grzeszczyk -- Authors contributed equally.}
\begin{abstract}

This study examines the effect of COVID-19 pandemic and associated lockdowns on access to crucial diagnostic procedures for breast cancer patients, including screenings and treatments. To quantify the impact of the lockdowns on patient outcomes and cost, the study employs a mathematical model of breast cancer progression. The model includes ten different  states that represent various stages of health and disease, along with the four different stages of cancer that can be diagnosed or undiagnosed.  
% Probabilities of transitions between states were used from literature or found empirically. 
The study employs a natural history stochastic model to simulate the progression of breast cancer in patients. The model includes transition probabilities between states, estimated using both literature and empirical data. The study utilized a Markov Chain Monte Carlo simulation to model the natural history of each simulated patient over a seven-year period from 2019 to 2025. The simulation was repeated 100 times to estimate the variance in outcome variables. The study found that the COVID-19 pandemic and associated lockdowns caused a significant increase in breast cancer costs, with an average rise of 172.5 million PLN (95\% CI [82.4, 262.6]) and an additional 1005 breast cancer deaths (95\% CI [426, 1584]) in Poland during the simulated period. While these results are preliminary, they highlight the potential harmful impact of lockdowns on breast cancer treatment outcomes and costs.

\end{abstract}

\section{Introduction}
The COVID-19 pandemic impacted the lives of people around the world. To slow down the spread of the disease, many countries introduced lockdown restrictions in form of banning gatherings, limiting outdoor trips and canceling public events \cite{koh2020covid}. While lockdowns positively influenced the pandemic progression (decreased doubling time) \cite{lau2020positive} or even environment \cite{arora2020coronavirus}, the negative impact on mental health \cite{adams2022impact}, physical fitness \cite{tsoukos2022effects}, dietary habits \cite{bennett2021impact} and other important aspects of our lives are evident. In this work we analyze the effect of pandemic lockdowns on breast cancer care in Poland. 
%After the pandemic ended, research is conducted to analyse the pros and cons of undertaken strategies.

Breast cancer is the most frequent cause of cancer deaths among women \cite{torre2015global} and is a high burden to public finance. There is an estimated 2.3 million women diagnosed with breast cancer and 685,000 deaths globally in 2020 \cite{sung2021global}. The direct cause of breast cancer is unknown, but there exist a number of risk factors like obesity, late menopause or alcohol use \cite{key2001epidemiology}. Since there are few to no symptoms at the early stage of breast cancer, many countries introduced screening programs in the form of free mammography procedures to support the early detection of the disease \cite{tabar2003mammography}. COVID lockdowns resulted in restricted access to healthcare \cite{goyal2021restricted} which consequently reduced the number of diagnosed and treated breast cancer patients. 

In this paper, we present a Markov Model-based approach to the Monte Carlo simulation of breast cancer disease progression. The nodes of the model are different states or cancer stages that the subject can be in at a specific point in time. The probabilities of transitions between states are computed based on the existing literature and empirical experiments. We use this method to conduct 100 repetitions of seven-year-long simulations on 1\% of the total women population in Poland. In the simulation, we consider the direct costs (medicines, surgeries, chemotherapy), indirect costs (premature death, absenteeism, disability) of breast cancer and statistics of the number of subjects in all states. We conduct two types of experiments. First, we perform the simulation taking into consideration the impact of COVID lockdowns on the accessibility of public healthcare, screening programs and treatment. Secondly, we conduct the simulation as if there was no pandemic. We extrapolate results to the population of the entire country.

The main contributions of this paper are as follows:
\begin{enumerate}
  \item We present a Markov Model-based simulation of the progression of breast cancer.
  \item We analyze the impact of COVID lockdowns on mortality and healthcare costs using a comparison of simulations conducted on the population of Polish women with and without the simulated effect of pandemic. 
  \item We provide a publicly available code to simulate the progression of breast cancer: \url{https://github.com/SanoScience/BC-MM}.
\end{enumerate}

The rest of the paper is structured as follows. In Section~\ref{sec:related_work}, we describe the existing methods for the simulation of disease progression and present our approach to breast cancer modeling based on Markov Models in Section \ref{sec:methodology}. We show the results of simulations with and without the effects of pandemic and discuss how COVID-19 impacted breast cancer patients and the costs of the disease in Section~\ref{sec:results} and conclude in Section~\ref{sec:conclusion}.
\section{Related work}
\label{sec:related_work}

In this section, we describe works presented in the literature related to the investigation of the impact of the COVID-19 pandemic on healthcare, modeling the progression of diseases and analysis of disease costs in public finance.

\subsection{The impact of COVID-19 pandemic on healthcare}
Since the beginning of the pandemic, researchers have been concerned about the possible, negative side effects of lockdowns \cite{schippers2020greater}. Paltrinieri \textit{et al.} \cite{paltrinieri2021beyond} reported that 35.1\% lifestyle survey participants encountered worsening of physical activity during lockdowns. Similar concerns were presented by Tsoukos and Bogdanis \cite{tsoukos2022effects} who described lower body fitness, poorer agility tests results and increased body mass index in adolescent students as the effects of a five-month lockdown. The negative impact of lockdowns does not end on the deterioration of physical fitness. Mental health is one of the factors that suffered the worst during the pandemic. Adams \textit{et al.} \cite{adams2022impact} discussed a significant decrease in self-reported mental health in the United States. The self-harm incidents due to stress related to COVID-19 were reported in India \cite{sahoo2020self}. Cases of depression, anxiety and post-traumatic stress disorders were expected to rise in Sub-Saharan Africa \cite{semo2020mental}.

During the pandemic, access to healthcare, especially related to the treatment of other diseases was limited \cite{goyal2021restricted}. Many global healthcare resources were reallocated to prevent and treat coronavirus infections \cite{chudasama2020impact}. The expected results of the depletion of healthcare resources were the increase of COVID-19 and all-cause mortality \cite{randolph2020herd}. Additionally, more than 28 million surgeries were expected to be canceled in the UK due to lockdowns \cite{covidsurg}. In most cases, those were operations for benign diseases, however, the effect cannot be neglected. Jiang \textit{et. al} \cite{jiang2021more} described examples of the co-epidemics of COVID-19 and other infectious diseases and potential negative effects on the treatment of non-communicable and chronic diseases.

Concerns regarding the impact of COVID-19 on the treatment of diseases give a justified basis for the analysis the influence of lockdowns on breast cancer prevalence and costs. As reported by Gathani \textit{et al.} \cite{gathani2021covid}, there was a steep decrease in the number of referrals for suspected breast cancer (28\% lower) and breast cancer diagnosis (16\% lower) in the UK in 2020. Yin \textit{et al.} \cite{yin2020breast} describe the decline in the usage number of 3 services (breast imaging, breast surgery and genetics consultation) in the early stages of the pandemic. In \cite{figueroa2021impact}, the pauses in screening programs that occurred in various countries were described, and disease modeling was mentioned as one of the possible approaches to analyse the repercussions of COVID-19 on breast cancer.  In this paper, we analyse the impact of those radical changes in breast cancer diagnosis and treatment on the costs of breast cancer in public finance.

\subsection{Modelling progression of the disease}
There are multiple methods for developing disease models for the purpose of conducting simulations \cite{kopec2010validation}. One of the approaches to stochastic process modeling (like the progression of the chronic disease) and economic impact analysis is the utilization of Markov Modelling \cite{briggs1998introduction}. In such a graph model, nodes represent stages of the disease and edges the probabilities of moving from one state to another. For instance, Liu \textit{et al.} \cite{liu2015efficient} presented the Continuous-Time Hidden Markov Model for Alzheimer's disease progression. In \cite{ramezankhani2018diabetes}, a multi-state semi-Markov model was used to investigate the impact of type 2 diabetes on the co-occurrence of cardiovascular diseases.

Markov Model can be successfully utilized to conduct an analysis of breast cancer. Momenzadeh \textit{et al.} \cite{momenzadeh2020using} used hidden Markov Model to predict the recurrence of breast cancer, while Pobiruchin \textit{et al.} \cite{pobiruchin2016method} presented a method for Markov Model derivation out of real-world datasets (cancer registry's database). Breast cancer modeling was also used to investigate the decline in screening, delays in diagnosis and delays in treatment during COVID-19 pandemic in the USA \cite{alagoz2021impact}. Alagoz \textit{et al.} \cite{alagoz2021impact} developed three models representing each issue. The conducted simulation exposed that there is a projected excess of breast cancer deaths by 2030 due to the pandemic. In this paper, we present the results of the experiments conducted with Monte Carlo simulation based on the Markov Model of breast cancer progression in Poland. 

\subsection{Costs of breast cancer care}
The analysis of disease costs for public finance is a difficult task as there are different methods that could be used and various types of costs that have to be taken into consideration \cite{campbell2009costs}. Costs in pharmacoeconomics can be divided into four categories: direct medical costs, direct non-medical costs, indirect costs and intangible costs \cite{petryszyn2014non}. Direct medical costs are the easiest to determine. They include the costs of medicines, diagnostic tests, hospital visits etc. Direct non-medical costs are costs mainly related to the treatment of the patient, but not having a medical basis. Another group of costs are indirect costs. They are mainly related to the loss of productivity associated with the patient's illness or death. The intangible costs are the costs associated with pain, suffering, fatigue and anxiety associated with the disease, as well as side effects of treatment such as nausea. They are difficult to estimate and measure, as they are mainly related to the patient's feelings \cite{rascati2013essentials}. In this paper, we take into consideration direct and indirect costs only. 

Depending on the methodology the calculated costs may highly vary (e.g. \$US20,000 to \$US100,000 of the per-patient cost) \cite{campbell2009costs}. Different studies analyse different types of costs, making it difficult to compare them. In \cite{baker1998use}, the mathematical model of continuous tumor growth and screening strategies was applied for the evaluation of screening policies. In \cite{elixhauser1991costs}, cost-effectiveness studies were used to estimate the costs of breast cancer screening per year of life saved to be \$13,200-\$28,000. The total cost of breast cancer in the USA was estimated to be \$3.8 billion. Blumen \textit{et al.} \cite{blumen2016comparison} conducted a retrospective study to compare the treatment costs by tumor stage and type of service. The analysis was undertaken on a population selected from the commercial claims database which facilitated the study as costs-related data was directly available.

\section{Methodology}
\label{sec:methodology}

\begin{figure}[t]
    \centering
    \includegraphics[width=\textwidth]{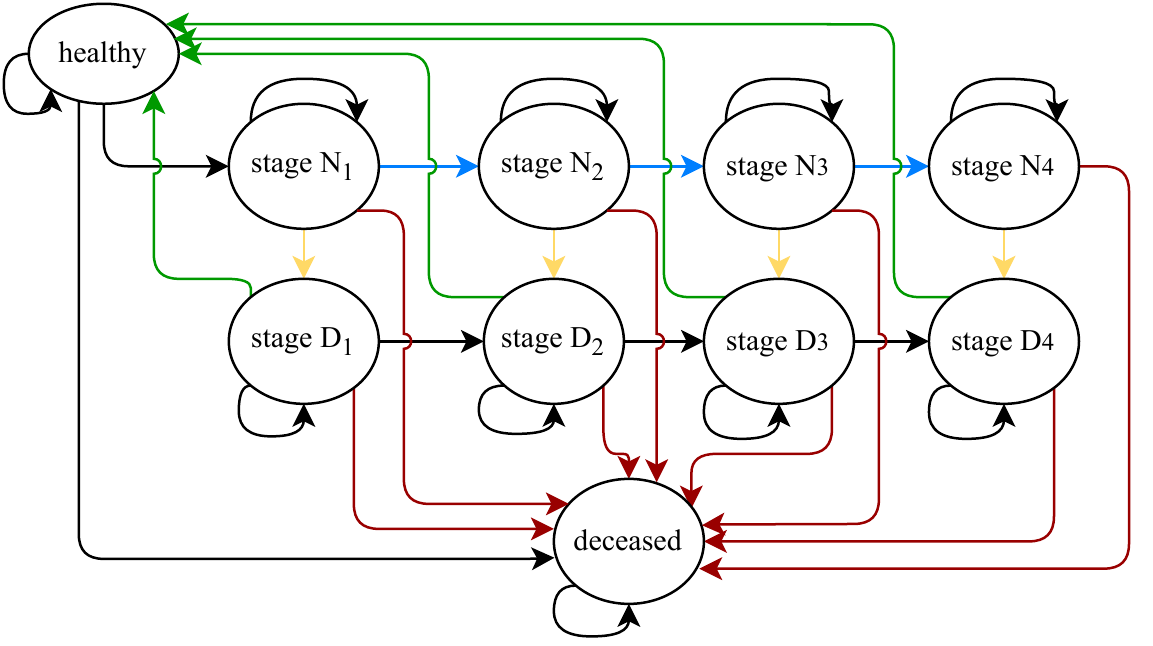}
    \caption{Breast cancer Markov Model with 10 states. Blue arrows indicate transitions between different stages of breast cancer, yellow ones show the diagnosis of breast cancer, green arrows indicate that the patient was healed and red represent the death event related to breast cancer.}
    \label{fig:bcmm}
\end{figure}

In this section, we describe the Markov Model used for the simulation of breast cancer progression. We define the types and values of costs used  and provide details on the parameters used in simulations. For more details on the actual algorithmic implementation refer to the source code available at \url{https://github.com/SanoScience/BC-MM}.

\subsubsection{Breast cancer Markov Model}
We use the Monte Carlo simulation based on the Markov Model to describe the course of breast cancer disease. The time horizon of the analysis is divided into equal time increments, called Markov cycles. During each cycle, the patient can transition from one state to another. Arrows connecting two different states indicate allowed transitions. We applied values of those transitions using clinical information, derived from previous studies or estimated empirically. The arrows from the state to itself indicate that the patient may remain in that state during cycles \cite{sonnenberg1993markov}. Transition probabilities are used to estimate the proportion of patients who will transfer from one state to another. The probability of an event occurring at a constant rate ($r$) during time ($t$) can be expressed by the equation:
\begin{equation}
    p=1-e^{-rt}
\end{equation}

In Figure \ref{fig:bcmm}, we present the Markov Model describing the progression of breast cancer. There are ten states in our model: \textit{healthy}, four states describing a non-diagnosed person with breast cancer at four stages \cite{amin2017ajcc} of the disease ($stage N_{i}$, where $i \in \{1, 2, 3, 4\}$), four states for diagnosed stages ($stage D_{i}$, where $i \in \{1, 2, 3, 4\}$) and \textit{deceased}. \textit{Deceased} is a terminal stage of our model and if a subject reaches this state its simulation is terminated. We follow The American Joint Committee on Cancer which defines four stages of breast cancer \cite{amin2017ajcc}.

\subsubsection{Simulation}

We assume that each Markov cycle is equal to one week.  Breast cancer is a disease that develops for many years, however, the longer the simulated period is, the less reliable are the results due to assumptions made about the future. Therefore, we set the number of cycles to 364, corresponding to 7 years (assuming that every year has 52 weeks). This period allows us to measure the long-term effects of the COVID-19 pandemic. We set the beginning of the simulation to January 1st 2019 so that the simulation  stabilizes (reaches equilibrium) before the 2020 year with a pandemic. We conduct two types of simulations, one taking into account COVID-19 lockdowns, and one which assumes that there was no effect of COVID-19 on breast cancer treatment and progression. We assume that lockdowns in Poland were lasting from March 2020 till the beginning of March 2021. We repeat  100 times each type and average the collected results.

In the simulation, we take into account malignant breast cancer only (C50 ICD-10 code). Stage 0 of breast cancer (which has a different ICD-10 code) often has a 100\% 5-year survival rate. Thus, we omit this stage in the analysis as it should not have a significant impact on costs and survival. We simulate the breast cancer progression in women as this sex accounts for most of the cases of the disease. We conduct computation on 1\% of the representative women population in Poland with the age distribution according to Table \ref{tab:age_distribution} - after the end of simulations, we multiply results by 100. The minimum age of simulated patients is set to 25 because below this age the occurrence of breast cancer is rare (Table \ref{tab:2019diag}). We increase the age of each simulated person every 52 cycles.

To find the number of women diagnosed with breast cancer in 2019 we compute a linear trend line based on the available data. There were 143,911, 151,831, 158,534, 166,031, and 174,005 patients with breast cancer in 2010, 2011, 2012, 2013, and 2014 respectively \cite{nojszewska2016ocena}. According to Agency for Health Technology Assessment and Tariff System in Poland, those numbers rose to 227,784 and 242,838 in 2016 and 2017 \cite{aotm}. The projected trend line indicated that in 2018 and 2019 there were 247,013 and 263,590 women with this disease in Poland. Taking into consideration the distribution of cancer stages among women with the diagnosed disease in the UK \cite{CRUK} and the distribution of age (Table \ref{tab:age_distribution}) we derive the number of diagnosed women in Poland by cancer stage and by age (Table \ref{tab:diag_undiag}). In addition, we estimate the number of undiagnosed people. We assume that breast cancer would only be detected using mammography and follow-up diagnostic regimen, and around 71\% of patients show up at this procedure \cite{NHS}. In 2019 the number of mammography tests was 1.041 million with 19620 cases detected (2\%). The number of people who fell ill in 2019 was 263,590. This is 2\% of the 71\% of people who appeared on mammograms. On this basis, the remaining people who did not show up on the mammogram and would have a positive result can be calculated. They are 2\% of the remaining 29\% that should come for a mammogram. The estimated number of people is 108,427. Using this number and information about the percentage of patients in a specific stage, we calculate the number of undiagnosed patients in stages II, III and IV (Table \ref{tab:diag_undiag}). The same strategy for stage I destabilizes the model. Thus, we set the number of undiagnosed patients in the first stage to the same value as for those diagnosed in the first stage in 2019. We make this assumption due to the fact that people in stage I very often do not have symptoms yet.

\begin{table}[t]
    \centering
    \caption{The age distribution of Polish women above 25 years in 2019 \cite{gus}.}
    \label{tab:age_distribution}
    \begin{tabular}{c|r|r}
        Age & Number of women & Percentage \\ \hline
        25-29 & 1,233,777\hspace{5mm}  & 8\% \\ 
        30-34 & 1,436,161\hspace{5mm}  & 10\% \\ 
        35-39 & 1,596,757\hspace{5mm}  & 11\% \\ 
        40-44 & 1,502,164\hspace{5mm}  & 10\% \\ 
        45-49 & 1,294,636\hspace{5mm}  & 9\% \\ 
        50-54 & 1,142,203\hspace{5mm}  & 8\% \\ 
        55-59 & 1,234,478\hspace{5mm}  & 8\% \\ 
        60-64 & 1,460,924\hspace{5mm}  & 10\% \\
        65-69 & 1,359,815\hspace{5mm}  & 9\% \\ 
        70-74 & 1,013,368\hspace{5mm}  & 7\% \\ 
        75-79 & 640,118\hspace{5mm}    & 4\% \\ 
        80-84 & 581,529\hspace{5mm}    & 4\% \\ 
        85+   & 583,545\hspace{5mm}    & 4\% \\ \hline
        Total & 15,079,475\hspace{5mm} & 100\% \\ 
    \end{tabular}
\end{table}

\begin{table}[t]
    \centering
    \caption{The distribution of patients diagnosed with breast cancer in 2019 in Poland \cite{rejestr}.}
    \label{tab:2019diag}
    \begin{tabular}{c|c|c|c|c|c|c|c|c|c|c|c|c|c|c}
        Age & 0-24 & 25-29 & 30-34 & 35-39 & 40-44 & 45-49 & 50-54 & 55-59 & 60-64 & 65-69 & 70-74 & 75-79 & 80-84 & 85+ \\ \hline
        \# & 8 & 79 & 292 & 697 & 1,252 & 1,591 & 1,832 & 2,120 & 2,970 & 3,377 & 2,039 & 1,442 & 1,107 & 814 \\ \hline
        \% & 0.0 & 0.4 & 1.5 & 3.6 & 6.4 & 8.1 & 9.3 & 10.8 & 15.1 & 17.2 & 10.4 & 7.3 & 5.6 & 4.1 \\ 
    \end{tabular}
\end{table}

\begin{table}[t]
    \centering
    \caption{The projected distribution of diagnosed and undiagnosed women with breast cancer in 2019 in Poland.}
    \label{tab:diag_undiag}
    \begin{tabular}{c|rrrr|rrrr}
        ~ & \multicolumn{4}{c|}{Diagnosed} & \multicolumn{4}{c}{Undiagnosed} \\ \hline
        Age \textbackslash Stage & I\hspace{2mm} & II\hspace{2mm} & III\hspace{2mm} & IV\hspace{2mm} & I\hspace{2mm} & II\hspace{2mm} & III\hspace{2mm} & IV\hspace{2mm} \\ \hline
        25-29 & 461\hspace{2mm} & 441\hspace{2mm} & 102\hspace{2mm} & 57\hspace{2mm} & 461\hspace{2mm} & 181\hspace{2mm} & 42\hspace{2mm} & 24\hspace{2mm} \\ 
        30-34 & 1,705\hspace{2mm} & 1,630\hspace{2mm} & 377\hspace{2mm} & 211\hspace{2mm} & 1,705\hspace{2mm} & 670\hspace{2mm} & 155\hspace{2mm} & 87\hspace{2mm} \\ 
        35-39 & 4,069\hspace{2mm} & 3,891\hspace{2mm} & 900\hspace{2mm} & 504\hspace{2mm} & 4,069\hspace{2mm} & 1,600\hspace{2mm} & 370\hspace{2mm} & 208\hspace{2mm} \\ 
        40-44 & 7,308\hspace{2mm} & 6,989\hspace{2mm} & 1,617\hspace{2mm} & 906\hspace{2mm} & 7,308\hspace{2mm} & 2,875\hspace{2mm} & 665\hspace{2mm} & 373\hspace{2mm} \\ 
        45-49 & 9,287\hspace{2mm} & 8,881\hspace{2mm} & 2,055\hspace{2mm} & 1,151\hspace{2mm} & 9,287\hspace{2mm} & 3,653\hspace{2mm} & 845\hspace{2mm} & 474\hspace{2mm} \\ 
        50-54 & 10,694\hspace{2mm} & 10,227\hspace{2mm} & 2,366\hspace{2mm} & 1,326\hspace{2mm} & 10,694\hspace{2mm} & 4,207\hspace{2mm} & 973\hspace{2mm} & 545\hspace{2mm} \\ 
        55-59 & 12,375\hspace{2mm} & 11,834\hspace{2mm} & 2,738\hspace{2mm} & 1,534\hspace{2mm} & 12,375\hspace{2mm} & 4,868\hspace{2mm} & 1,126\hspace{2mm} & 631\hspace{2mm} \\ 
        60-64 & 17,337\hspace{2mm} & 16,579\hspace{2mm} & 3,836\hspace{2mm} & 2,150\hspace{2mm} & 17,337\hspace{2mm} & 6,820\hspace{2mm} & 1,578\hspace{2mm} & 884\hspace{2mm} \\ 
        65-69 & 19,713\hspace{2mm} & 18,851\hspace{2mm} & 4,361\hspace{2mm} & 2,444\hspace{2mm} & 19,713\hspace{2mm} & 7,754\hspace{2mm} & 1,794\hspace{2mm} & 1,005\hspace{2mm} \\ 
        70-74 & 11,902\hspace{2mm} & 11,382\hspace{2mm} & 2,633\hspace{2mm} & 1,476\hspace{2mm} & 11,902\hspace{2mm} & 4,682\hspace{2mm} & 1,083\hspace{2mm} & 607\hspace{2mm} \\ 
        75-79 & 8,417\hspace{2mm} & 8,050\hspace{2mm} & 1,862\hspace{2mm} & 1,044\hspace{2mm} & 8,417\hspace{2mm} & 3,311\hspace{2mm} & 766\hspace{2mm} & 429\hspace{2mm} \\ 
        80-84 & 6,462\hspace{2mm} & 6,179\hspace{2mm} & 1,430\hspace{2mm} & 801\hspace{2mm} & 6,462\hspace{2mm} & 2,542\hspace{2mm} & 588\hspace{2mm} & 330\hspace{2mm} \\ 
        85+ & 4,752\hspace{2mm} & 4,544\hspace{2mm} & 1,051\hspace{2mm} & 589\hspace{2mm} & 4,752\hspace{2mm} & 1,869\hspace{2mm} & 432\hspace{2mm} & 242\hspace{2mm} \\ 
    \end{tabular}
\end{table}

\subsubsection{State transition probabilities}
We derive the following state transition probabilities:
\begin{enumerate}
    \item $P(healthy \rightarrow stage N_{1})$ - the probability of developing breast cancer,
    \item $P(healthy \rightarrow deceased)$ - the probability of non-cancer related death,
    \item $P(stage N_{i} \rightarrow stage N_{i+1})$ - the probability of cancer stage increase,
    \item $P(stage N_{i} \rightarrow stage D_{i})$ - the probability of cancer diagnosis,
    \item $P(stage N_{i} \rightarrow deceased)$ - the probability of breast cancer death,
    \item $P(stage D_{i} \rightarrow stage D_{i+1})$ - the probability of cancer stage increase,
    \item $P(stage D_{i} \rightarrow healthy)$ - the probability of healing,
    \item $P(stage D_{i} \rightarrow deceased)$ - the probability of breast cancer death.
\end{enumerate}

To simulate the effects of covid lockdowns we modify three transition probabilities.  The probability of cancer diagnosis is decreased because of lockdowns and restricted access to healthcare. The probability of breast cancer-related death is increased due to a lack of proper healthcare assistance, and the probability of healing is decreased due to poorer treatment during the COVID-19 pandemic. Numerically we implement the models as follows.

We assume  \textbf{probability of developing breast cancer} is only dependent on women's age. We set this probability in the following manner - age (probability; probability in one cycle): 20 (0.1\%; 0.0002\%), 30 (0.5\%; 0.001\%), 40 (1.5\%; 0.0029\%), 50 (2.4\%; 0.0046\%), 60 (3.5\%; 0.0067\%), 70 (4.1\%; 0.0079\%), 80 (3.0\%; 0.0058\%) \cite{bc_facts}. We define the \textbf{probability of non-cancer related death} according to the life tables from 2019 \cite{gus}. There are multiple resources defining the Progression-Free Survival parameter which is a time during which the patient lives with the disease but their state is not worsening. For example, Haba-Rodriguez \textit{et al.} \cite{de2008time} state that the median of PFS varies between 4 to 18 months. Thus, we empirically set the \textbf{probability of cancer stage increase for diagnosed patient} to $p=k(1-e^{- \lambda t})$ where $k$ is 0.0009, $t$ is the number of cycles and $\lambda$ is 10, 15, 20 or 25 depending on the stage of the disease. It is difficult and highly uncertain to assess the progression of breast cancer in an undiagnosed patient as no data exist that describe those transitions. Therefore, we define the \textbf{probability of cancer stage increase for undiagnosed women} in the same manner as in the case of diagnosed cases and set $\lambda$ to 20, 25, 30 or 35 depending on the stage which is a reasonable approximation.  

We define the \textbf{probability of healing} based on the 5-year Disease Free Survival (DFS) parameter which changes depending on the cancer stage \cite{nowikiewicz2015overall}. The 5-year DFS in 2019 was 0.987 (stage I), 0.873 (stage II), 0.52 (stage III), and 0.037 (stage IV). We decrease those values during lockdowns by 19\% (due to a 19\% decrease in hospitalizations \cite{nfz}) to 0.801 (stage I), 0.708 (stage II), 0.422 (stage III), and 0.03 (stage IV) and set the probability of healing to $p=k(1-e^{- \lambda t})$ where $k$ is set to $\frac{1}{3}$ and $\lambda$ is computed based on the 5-year DFS. The \textbf{probability of death for diagnosed patient} is computed from the 5-year survival rate which indirectly provides the probability of death within 5 years. Taking into consideration the 5-year survival in stages I, II, III and IV of 0.975, 0.856, 0.44, 0.23 \cite{nowikiewicz2015overall}, we compute $\lambda$ parameter of the probability of death in cycle $\leq t$ ($p(x \leq t)=1-e^{- \lambda t}$) to be 0.0061, 0.0302, 0.1642 and 0.2939 respectively. For covid simulation according to predictions that the mortality rate might increase by 9.6\% \cite{maringe2020impact} the $\lambda$ parameter is set to 0.0056, 0.0344, 0.1903, 0.3715 for every stage. The 3-month delay in cancer diagnosis may increase the chances of premature death by 12\% \cite{bish2005understanding}. We, therefore, find the \textbf{probability of death for undiagnosed patient} by increasing the 5-year death probability for diagnosed patients and compute $\lambda$ for those probabilities equal to 0.0061, 0.0349, 0.1989, 0.3932 depending on the stage. 

In 2019, approximately 7\% of all women aged 25 and over had a mammogram. The situation changed in 2020 when the number of mammograms decreased by over 305,000, which was a decrease of about 29\%. We assume that malignant breast cancer can only be detected by mammography and diagnostic follow-ups. The newly detected cases in 2019 (19,620) accounted for 2\% of all mammograms performed. In 2020, the number of detected cases decreased due to the COVID-19 pandemic. The average annual growth rate of new cases of breast cancer is 3\%. This means that around 20,209 new cases of cancer should have been diagnosed in 2020. We assume that the percentage of positive mammograms did not change, which will bring the number of detected cases to about 13,873. This is a difference of about 6,000 cases compared to what should have been detected. About 12.5\% of mammograms are thought to have a false-negative result and data shows that only 71\% of all women show up for the examination. In 2020, the number of these women probably decreased even more (by 29\%). Therefore, we define the \textbf{probability of diagnosis} in one year as $p=\frac{l_{pos}}{l_{pos}+l_{fneg} +l_{nm}}$ where $l_{pos}$ is the number of positive mammography cases, $l_{fneg}$ is the number of false negative mammography and $l_{nm}$ is the number of women that should have taken part in the mammography and would have had positive results. Thus, this probability for 2019 is 12.4\% and 11.6\% during lockdowns.

\subsubsection{Costs of breast cancer in Poland}
We collect two types of costs during simulation: direct costs and indirect costs. We divide the latter into indirect costs related to premature death and other indirect costs related to absenteeism, presenteeism, absenteeism of informal caregivers, presenteeism of informal carers, disability etc.

We derive direct per-person, per-stage costs in the following way. We estimate the total direct costs for 2019 based on the estimated number of breast cancer patients and direct costs in 2010-2014 years \cite{nojszewska2016ocena} to be 846,653 thousand PLN. We follow the distribution of the stage-specific costs in \cite{mittmann2014health}. We compute that direct per-stage yearly costs, based on the number of patients in every stage, in the simulation are: stage I (25\% of total costs, 207,560 thousand PLN, 1881 PLN per person), stage II (38\%, 325,108 thousand PLN, 3,185 PLN per person), stage III (25\%, 215,076 thousand PLN, 5,573 PLN per person), stage IV (12\%, 98,909 thousand PLN, 7,869 PLN per person). We add those costs (divided by the number of cycles in one year) for every diagnosed woman in each cycle of the simulation.

We compute indirect death costs by averaging the per-person death costs related to breast cancer patients in 2010-2014 years \cite{nojszewska2016ocena}. The average value of premature death cost is 123,564 PLN. We add this value every time a simulated person enters \textit{deceased} state from one of the breast cancer stages. We estimate other indirect costs in the same way. The average value of indirect per-patient cost in 2010-2014 years is 13,159 PLN. We add this value (divided by the number of cycles in year) for every patient in the $stage D_{i}$ state in every cycle.

\subsubsection{Experimental setup}
We develop the Monte Carlo simulation with Python 3.10 programming language. We conduct all simulations on a 1.80 GHz Intel Core i7 CPU and 16 GB RAM. The execution of all simulations took 3 hours to complete.

\section{Results and discussion}
\label{sec:results}

\subsubsection{Costs of breast cancer} In Table \ref{tab:detailed_results}, we present changes in average direct and indirect costs over the 7-years period. In all cases, the total costs incurred in the absence of lockdowns are smaller than the ones that resulted from the simulation with the COVID-19 pandemic. However, the only statistically significant difference (p-value < 0.001) is in the case of indirect costs related to premature death. This is reasonable because breast cancer is a long-lasting disease and the costs of treatment or patient care did not change drastically due to lockdowns. On the other hand, delayed diagnoses and surgeries resulted in more premature deaths. The impact of the pandemic is also reflected in the total costs of breast cancer (Table \ref{tab:total_results}). the pandemic resulted in an increase in breast cancer costs of 172.5 million PLN (average total costs with covid - average total costs without covid) with 95\% confidence interval (CI) of [82.4, 262.6]. The difference between total costs with and without lockdowns is statistically significant (p-value < 0.001). The positive influence of lockdowns on the progression of the pandemic should not be neglected. However, the presented results suggest that lockdowns had a negative impact on overall disease treatment, both socially and economically.

\begin{table}[t]
    \centering
    \caption{Direct and indirect simulated costs (in million PLN) of breast cancer in Poland with and without COVID-19 pandemic.}
    \label{tab:detailed_results} 
%    \begin{tabular}{c|cc|cc|cc}
    \begin{tabular}{c D{,}{\pm}{4.4} D{,}{\pm}{-1} | D{,}{\pm}{-1} D{,}{\pm}{-1} | D{,}{\pm}{-1} D{,}{\pm}{-1} }
              & \multicolumn{2}{c|}{DIRECT} &                    \multicolumn{2}{c|}{INDIRECT\_DEATH} & \multicolumn{2}{c}{INDIRECT\_OTHER}   \\
        year  &   \multicolumn{1}{c}{NO COVID}                    &    \multicolumn{1}{c|}{COVID}    &     \multicolumn{1}{c}{NO COVID}     & \multicolumn{1}{c|}{COVID}             &     \multicolumn{1}{c}{NO COVID}     & \multicolumn{1}{c}{COVID} \\ \hline
        2019     &    763,4          &    764,4 & 625,79   & 625,93   & 3297,14   & 3302,13 \\
        2020     &    770,7          &    771,7 & 827,93   & 875,106  & 3317,26   & 3321,28 \\
        2021     &    775,10         &    775,9 & 990,107  & 996,113  & 3337,36   & 3338,34 \\ 
        2022     &   777,11          &   777,12 & 1087,118 & 1119,120 & 3352,40   & 3355,44 \\
        2023     &   777,11          &   779,13 & 1186,116 & 1197,121 & 3367,43   & 3372,50 \\
        2024     &   775,12          &   779,14 & 1275,120 & 1270,119 & 3377,49   & 3387,51 \\
        2025     &   774,13          &   777,15 & 1306,150 & 1340,30 & 3389,53   & 3399,56 \\ \hline
        TOTAL    &  5411,55          &  5422,64 & 7296,246 & 7420,267 & 23437,213 & 23474,236 \\

    \end{tabular}
\end{table}

\begin{table}[t]
    \centering
    \caption{Total simulated costs (in million PLN) of breast cancer in Poland.}
    \label{tab:total_results}
    \begin{tabular}{c| D{,}{\pm}{4} D{,}{\pm}{-1}}
              & \multicolumn{2}{c}{TOTAL} \\
        year  &   \multicolumn{1}{c}{NO COVID}                 &   \multicolumn{1}{c}{COVID}    \\ \hline
        2019    &    4686,77 & 4691,91 \\
        2020    &    4915,91 & 4967,107 \\
        2021    &   5101,116 & 5109,118 \\
        2022    &   5216,133 & 5251,122 \\
        2023    &   5329,122 & 5348,135 \\
        2024    &   5428,130 & 5435,133 \\
        2025    &   5469,170 & 5516,140 \\ \hline
        TOTAL&  36144,316 & 36317,333 \\

    \end{tabular}
\end{table}

\subsubsection{Breast cancer with and without COVID-19 pandemic} 
\begin{figure*}[t!]
\minipage{\textwidth}
  \includegraphics[width=\linewidth]{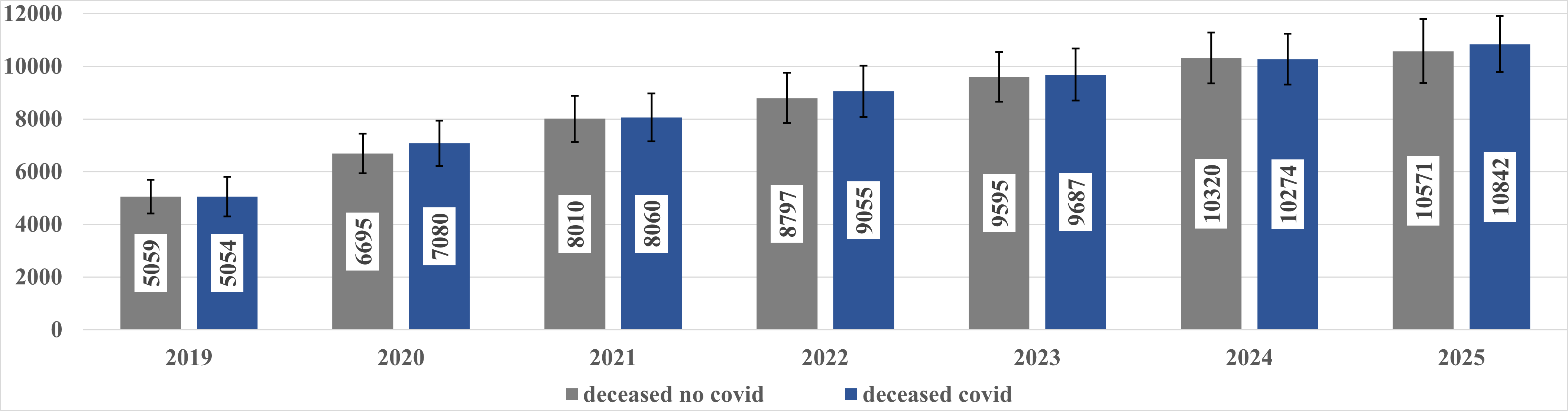}
\endminipage\hfill

\minipage{\textwidth}
  \includegraphics[width=\linewidth]{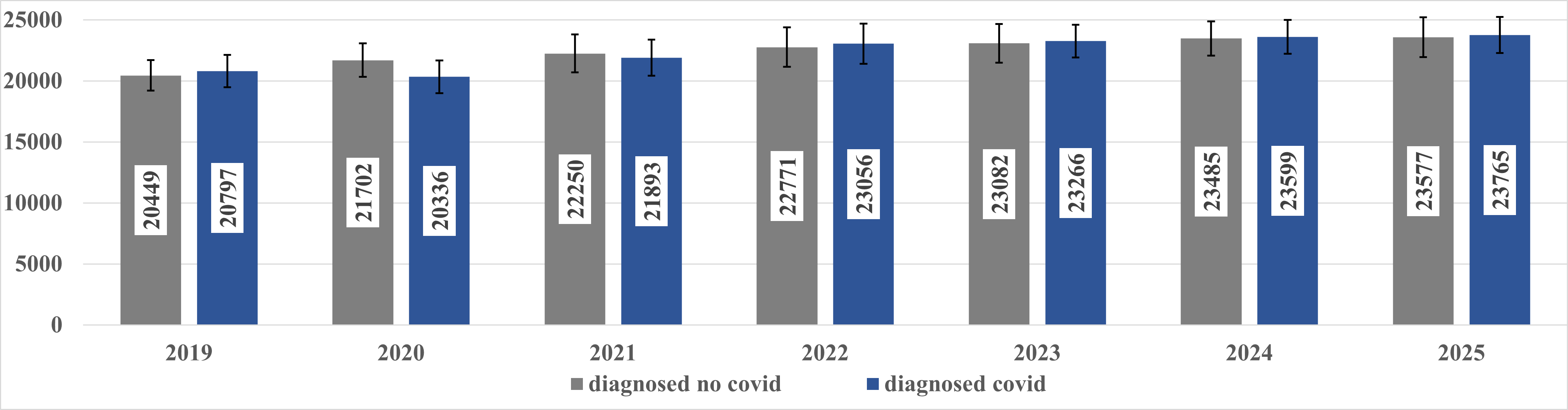}
\endminipage\hfill
\caption{The average number of breast cancer-related deaths (top) and the average number of breast cancer diagnoses (bottom) with and without lockdowns.}
\label{fig:charts}
\end{figure*}

Simulations also showed that there was a significant difference in the number of women deaths due to COVID-19. On average, during 7 years, 60,052 women died taking into consideration lockdowns. This number would be smaller by 1005 deaths if the pandemic did not occur (95\% CI [426, 1584]). Year-by-year visualization of deaths is presented in Figure \ref{fig:charts}. It can be noticed that delayed diagnoses and poorer treatment resulted in an overall increase in the number of deaths. The long-term effects will be visible in years to come. Figure \ref{fig:charts} depicts also the average number of cases of diagnosed breast cancer in Poland. There is a sharp decline in the number of diagnoses between covid and no covid simulations in 2020. The delayed diagnoses resulted in an increased probability of complications and death. In the following years, the trend was reversed and more of the delayed cases were diagnosed in covid simulation. However, the inefficient healthcare system is not capable of making up for the lost diagnostic time.

\subsubsection{Limitations} Our study is subject to limitations. First, the model and simulations presented are only approximations of the real-world situation, and therefore, the results should be interpreted with caution. Second, the impact of the COVID-19 pandemic on the costs associated with breast cancer is complex and rapidly evolving, and our study provides only a snapshot of the situation at the time of the analysis. Third, in order to build the model, we had to make several assumptions and rely on estimates or information from countries other than Poland due to a lack of national data. Access to healthcare and treatment may vary across different countries, and this may have resulted in overestimated or underestimated data in our model. Therefore, our findings should be interpreted in the context of these limitations and further research is needed to validate and expand our results. 
To account for the uncertainty around the course of the tumor, empirical fitting of transition probabilities was necessary. This is because, upon diagnosis, patients are immediately referred for treatment, leaving no research data on the disease's development. Furthermore, the study assumes that people with cancer did not directly die from coronavirus infection, but those at an advanced stage of the disease may have had a higher risk of succumbing faster after being infected with the pathogen. It is also worth noting that the model does not consider potential changes in healthcare policies or treatment protocols during the pandemic, which could have affected breast cancer care costs and patient outcomes. Despite these limitations, the study provides valuable insights into the potential impact of the pandemic on breast cancer care costs, and its findings could be beneficial to healthcare policymakers, clinicians, and researchers. Nevertheless, more research is necessary to confirm and expand the results presented in this study. 

\section{Conclusion}
\label{sec:conclusion}
In this study, we have used a Monte Carlo simulation approach and a Markov Model to analyze the effects of COVID-19 lockdowns on the costs and mortality of breast cancer in Poland. Our findings indicate a significant negative impact on breast cancer treatment, resulting in increased costs and higher mortality rates. Although these findings are preliminary, they offer important insights for future discussions on strategies that could be employed during future pandemics. As part of our ongoing research, we plan to conduct a sensitivity analysis of model parameters and expand our analysis to estimate the impacts of lockdowns on other diseases. 

\section*{Acknowledgements} This publication is partly supported by the European Union’s Horizon 2020 research and innovation programme under grant agreement Sano No. 857533 and the International Research Agendas programme of the Foundation for Polish Science, co-financed by the European Union under the European Regional Development Fund. We would like to thank Tadeusz Sat\l{}awa, Katarzyna Tabor, Karolina Tkaczuk, and Maja Wi\k{e}ckiewicz from Sano for help and their initial contribution in the development of the model. 

\bibliographystyle{splncs04}
\bibliography{bibliography}

\end{document}